\documentclass[sigconf]{acmart}
\AtBeginDocument{%
  \providecommand\BibTeX{{%
    \normalfont B\kern-0.5em{\scshape i\kern-0.25em b}\kern-0.8em\TeX}}}

\copyrightyear{2023}
\acmYear{2023}
\setcopyright{rightsretained}
\acmConference[ICER '23 V1]{Proceedings of the 2023 ACM Conference on
International Computing Education Research V.1}{August 7--11, 2023}{Chicago,
IL, USA}
\acmBooktitle{Proceedings of the 2023 ACM Conference on International
Computing Education Research V.1 (ICER '23 V1), August 7--11, 2023, Chicago,
IL, USA}
\acmDOI{10.1145/3568813.3600139}
\acmISBN{978-1-4503-9976-0/23/08}

\usepackage{listings}
\usepackage{tcolorbox}

\begin{document}
\lstset{frame=single}

\title{Exploring the Responses of Large Language Models to Beginner Programmers’ Help Requests}


\author{Arto Hellas}
\affiliation{%
  \institution{Aalto University}
  \country{Finland}}
\email{arto.hellas@aalto.fi}
\orcid{0000-0001-6502-209X}
  
\author{Juho Leinonen}
\affiliation{%
  \institution{The University of Auckland}
  \country{New Zealand}}
\email{juho.leinonen@auckland.ac.nz}
\orcid{0000-0001-6829-9449}

\author{Sami Sarsa}
\affiliation{%
  \institution{Aalto University}
  \country{Finland}}
\email{sami.sarsa@aalto.fi}
\orcid{0000-0002-7277-9282}

\author{Charles Koutcheme}
\affiliation{%
  \institution{Aalto University}
  \country{Finland}}
\email{charles.koutcheme@aalto.fi}
\orcid{0000-0002-2272-2763}

\author{Lilja Kujanpää}
\affiliation{%
  \institution{Aalto University}
  \country{Finland}}
\email{lilja.kujanpaa@aalto.fi}
\orcid{0009-0002-0205-7003}

\author{Juha Sorva}
\affiliation{%
  \institution{Aalto University}
  \country{Finland}}
\email{juha.sorva@aalto.fi}
\orcid{0009-0003-1727-1317}


\begin{abstract}
  
\textbf{Background and Context:} Over the past year, \emph{large language models} (LLMs) have taken the world by storm. In computing education, like in other walks of life, many opportunities and threats have emerged as a consequence. 

\noindent
\textbf{Objectives:} In this article, we explore such opportunities and threats in a specific area: responding to student programmers’ help requests. More specifically, we assess how good LLMs are at identifying issues in problematic code that students request help on. 

\noindent
\textbf{Method:} We collected a sample of help requests and code from an online programming course. We then prompted two different LLMs (OpenAI Codex and GPT-3.5) to identify and explain the issues in the students’ code and assessed the LLM-generated answers both quantitatively and qualitatively. 

\noindent
\textbf{Findings:} 
GPT-3.5 outperforms Codex in most respects. Both LLMs frequently find at least one actual issue in each student program (GPT-3.5 in 90\% of the cases). Neither LLM excels at finding \emph{all} the issues (GPT-3.5 finding them 57\% of the time). False positives are common (40\% chance for GPT-3.5). The advice that the LLMs provide on the issues is often sensible. The LLMs perform better on issues involving program logic rather than on output formatting. Model solutions are frequently provided even when the LLM is prompted not to. LLM responses to prompts in a non-English language are only slightly worse than responses to English prompts. 

\noindent
\textbf{Implications:} Our results continue to highlight the utility of LLMs in programming education. At the same time, the results highlight the unreliability of LLMs: LLMs make some of the same mistakes that students do, perhaps especially when formatting output as required by automated assessment systems. Our study informs teachers interested in using LLMs as well as future efforts to customize LLMs for the needs of programming education.

\end{abstract}





\maketitle

\section{Introduction}

Within the last year, large language models (LLMs) and tools built on them, such as ChatGPT and GitHub Copilot, have broken into the mainstream. Computing education research (CER), too, has seen an explosion of recent work exploring the opportunities and challenges that LLMs bring. Opportunities in computing education include the automation of natural-language explanations of code~\cite{macneil2022generating,sarsa2022automatic,macneil2023experiences,leinonen2023comparing}, personalized exercises~\cite{sarsa2022automatic,denny2022robosourcing}, enhanced error messages~\cite{leinonen2022using}, and assistance in solving CS1 exercises~\cite{wermelinger2023using}. Challenges include student over-reliance and plagiarism~\cite{finnieansley2022robots,becker2023programming,denny2023computing,prather2023its} as well as biases in generated content~\cite{becker2023programming,nadeem2020stereoset}.

For better or worse, vast numbers of students are already using LLMs to assist them in their studies. Such use is likely only to increase in the future. Some student use of LLMs will happen unofficially at each student’s discretion and will employ highly generic tools akin to ChatGPT\footnote{https://openai.com/blog/chatgpt} or programming-generic tools such as Codex\footnote{https://openai.com/blog/openai-codex}. The future may also see custom LLMs that have been designed to assist students of programming and that teachers adopt as official components of programming courses.

One potential application of LLMs is to respond to students’ help requests. In an ideal world, an LLM might assist a programming student who asks for help in many of the same ways that a good human teaching assistant would: the LLM might provide explanations and feedback, avoid falsehoods as well as instant “spoilers” about model solutions, foster conceptual understanding, challenge the student to reason about their work, adapt responses to the student’s current understanding, and in general promote learning. Such assistance might be provided rapidly and at scale.

We are not in that ideal world; LLMs are not pedagogical experts. In this work, we assess how LLMs respond to student help requests in the domain of introductory programming. Rather than dropping an LLM into an actual programming course and having students rely on it for assistance, we study a simulacrum of such a scenario: we take actual help requests collected during a programming course (and answered then by humans) and feed the requests as input to LLMs so that we the researchers may explore the responses. 

For us to characterize LLM responses to help requests in a particular context, we must be able to characterize those requests as well. Our first research question is therefore as follows:

\begin{itemize}
    \item[\textbf{RQ1}] \emph{When students in an introductory programming course request help, what sorts of issues are present in their code?}
\end{itemize}

\noindent This leads to our main question:

\begin{itemize}
    \item[\textbf{RQ2}] \emph{How do responses generated with large language models address the issues associated with students' help requests?} 
    \begin{itemize}
    \item[\textbf{(a)}] \emph{Are the responses thorough and accurate in identifying the issues in student code?}
    \item[\textbf{(b)}] \emph{Are there differences in response quality between prominent LLMs (ChatGPT-3.5 vs. Codex)?}
    \item[\textbf{(c)}] \emph{To what extent is response quality affected by prompting the LLM in a non-English language?}\footnote{The motivation for this subquestion is that, anecdotally, modern LLMs perform fairly well in various languages but best in English.}
    \item[\textbf{(d)}] \emph{What other themes of potential pedagogical relevance show up in the LLM responses (e.g., language style, presence of model solutions)?}
    \end{itemize}
\end{itemize}

\noindent The answers to these questions provide a picture of how well current LLMs perform in analyzing beginner students’ programs and commenting on them. Our findings also illustrate that there is still a ways to go if we are to reach the ideal sketched out above. On the other hand, the findings take the field a step closer to understanding how to use LLMs productively in computing education and, perhaps, closer also to designing custom LLMs for the needs of computing educators and students.




\section{Background}
\label{sec:background}

\subsection{Large Language Models}

Although large language models have only recently made a global breakthrough, the work that led to LLMs spans decades, drawing from advances in natural language processing and machine learning, as well as from increased availability of large quantities of data and computational resources. 

At their core, LLMs are deep learning models. They comprise of layers of vectors, where each cell (or ``neuron'') in a layer is a mathematical function that takes a vector as an input, has learnable parameters (or ``weights''), and produces an output as a weighted sum of the inputs.

A deep learning model is trained by providing training data to the network and adjusting the weights of the neurons so that the overall network learns to produce a desired output. Training requires large amounts of data, especially when the data is complex---for example, when sequential relations like word order are involved. For this reason, methods such as the \emph{long-short term memory recurrent neural network} (RNN)~\cite{hochreiter1997long} have emerged, which allow neurons to be connected with a directed graph that can represent a temporal sequence, and where the output of each neuron can be fed back to the network (in a recursion of sorts). The introduction of the \emph{attention} mechanism to RNN~\cite{bahdanau2014neural} enhanced the capture of long-range dependencies, leading to substantially improved performance on natural language processing. The attention mechanism further led to the \emph{transformer} architecture~\cite{vaswani2017attention}, which removed recurrent connections in favor of a \emph{self-attention} mechanism that improved the parallelization of training and reduced training time. 

The transformer architecture played a key role in the emergence of the \emph{generative pre-trained transformer} (GPT)~\cite{radford2018improving}. GPT was initially pre-trained (unsupervised learning) on a large data set in order for the model to infer fundamental rules such as grammar. This was followed by a fine-tuning phase, where the pre-trained model was further trained to handle various specific tasks such as classification, similarity detection, and so on. The original GPT had 117 million parameters (weights or neurons) and outperformed contemporary models on a number of  natural language processing benchmarks~\cite{radford2018improving}. Subsequent LLMs such as GPT-2~\cite{radford2019language}, GPT-3~\cite{brown2020language}, and InstructGPT~\cite{ouyang2022training} have built on these advances, increasing the number of parameters by several orders of magnitude 
and improving the fine-tuning process~\cite{radford2019language,brown2020language,ouyang2022training}.

Discussions about LLMs often feature humanizing phrases such as ``hallucination''~\cite{ji2022survey} or ``the AI thinks X.'' Nevertheless, and despite the dramatic advances, LLMs are at heart probabilistic models whose behavior is determined by data. Any output generated by an LLM is based on the input---the \emph{prompt}---and the previously learned parameters.

\subsection{Large Language Models in CER}

The emergence of large language models has sparked significant interest within CER, too~\cite{becker2023programming,macneil2023automatically}. Some of the initial studies focused on the performance of LLMs on introductory programming problems. For example, \citet{finnieansley2022robots} noted that the Codex LLM performed better than most introductory-level students, and similar observations were made in a data structures course as well~\cite{finnieansley2023my}; others have reported somewhat lower performance for GitHub Copilot, which is built on top of Codex~\cite{wermelinger2023using}. Researchers have also evaluated LLMs’ usefulness for creating new, personalized programming exercises~\cite{sarsa2022automatic} and explored ``robosourcing''~\cite{denny2022robosourcing}, where LLMs generate input for learnersourcing---that is, students take LLM-generated materials and improve on them. 

Another line of work in CER~\cite{sarsa2022automatic,macneil2022generating,macneil2023experiences,leinonen2023comparing} has looked at code explanations constructed by the Codex and GPT-3 LLMs, which have been optimized for source code and natural language, respectively.
Overall, LLMs have been found capable of explaining source code in natural language, which can be helpful for novices; there is some evidence that GPT-3 outperforms Codex~\cite{macneil2023experiences}, and that LLM-generated code explanations may be of higher quality than those created by students~\cite{leinonen2023comparing}. Recent work has also explored using Codex to explain and enhance error messages~\cite{leinonen2022using}.


Classroom evaluations are still relatively rare, as sufficiently performant LLMs emerged only very recently. Most research in CER has involved expert evaluations (e.g.,~\cite{sarsa2022automatic,leinonen2022using}) or lab studies (e.g.,~\cite{prather2023its}). A notable exception is the work of~\citet{macneil2023experiences}, who evaluated LLM-generated code explanations in an online course on web software development; another is the controlled study by~\citet{kazemitabaar2023studying}, where a group of novices with access to Codex outperformed a control group on code-authoring tasks.

As noted above, an LLM’s outputs are determined by prompts and the model’s parameters. Coming up with good inputs is key to generating meaningful output, so it makes sense that much of the LLM-based work in CER has involved some \emph{prompt engineering}. As an example, \citet{denny2022conversing} improved the performance of GitHub Copilot on introductory programming exercises from approximately 50\% to 80\% by exploring alternative prompts. Similarly, \citet{leinonen2022using} explored five different prompts for enhancing programming error messages and chose the prompt that lead to the best initial results. Prompt engineering may also involve a comparison of different LLMs~\cite{macneil2023experiences}. For a literature review on prompting (from a machine learning perspective), see~\citet{liu2023pre}.

To the best of our knowledge, there is no prior work on how LLMs perform on responding to help requests on programming problems---that is, scenarios where students have explicitly signaled that they require help.

%



\subsection{Novice Programmers and Errors}


Students learning to program are bound to face errors. In CER, early studies of novice errors focused on specific problems such as the ``Rainfall Problem''~\cite{soloway1982what,soloway1983cognitive,johnson1983bug,seppala2015we}. Later studies have evolved alongside new capabilities for data collection. Using data from automated assessment~\cite{ala2005survey,douce2005automatic,ihantola2010review,paiva2022automated} and programming environments that track students' process~\cite{ihantola2015educational}, researchers have quantified the types of errors that students face while programming~\cite{jadud2005first, denny2012all, vihavainen2014novices,dy2010detector,mccall2014meaningful}. Some errors are more frequent than others~\cite{spohrer1986novice}, some errors take more time to fix than others~\cite{denny2012all,brown2017novice,smith2019error,mccall2019new}, and the types of errors that students face tend to evolve~\cite{altadmri201537}. Data on errors informs teachers about the issues that their students frequently face, which does not always match the teachers' expectations~\cite{brown2017novice}.

Only some of the errors that students face are related to syntax, of course~\cite{altadmri201537,ettles2018common}; logic errors are also common, and varied. Ettles et al.~\cite{ettles2018common} sorted common logic errors in three categories: \emph{algorithmic errors} have a fundamentally flawed approach, \emph{misinterpretations} involve misinterpreting the task, and \emph{misconceptions} are flaws in programming knowledge. 
A related stream of research has sought to improve error messages, which when done right could lead to better learning~\cite{becker2019compiler,denny2021designing}, especially as regular error messages do not always match the underlying cause~\cite{mccall2014meaningful,becker2019compiler,dy2010detector}.

\section{Methodology}
\label{sec:methodology}

\begin{table*}[ht!]
\caption{Summaries of the exercises we analyzed. The `Count' column lists the number of help requests for each exercise. \label{tbl:programming-exercises-with-most-help-requests}}
\centering
\small
\vspace{-2mm}
\begin{tabular}{c l p{9cm}} 
 \toprule
  Count & Exercise name & Exercise description \\ 
 \midrule
    66 & Difference between two numbers                 & Writing a program that reads in two numbers and prints out their difference.  \\
    57 & Asking for a password                       & Creating a single-parameter function that takes in a password. Calling the function will repeatedly prompt for input until the user types in the password.   \\
    47 & Average of entered numbers       & Writing a program that reads in numbers from the user until the user types in 0. The program then prints out the average of the entered numbers or, if no numbers were entered, a specific string.  \\
    42 & Counting positive numbers     & Creating a function that counts the positive numbers in a given list. \\
    40 & Authentication                    & Writing a program that first asks for a username. If the username is ``admin,'' the program continues to ask for a password. The output of the program depends on whether the password was also correct. If the username is not ``admin,'' the program does not ask for a password and gives a specific output.  \\
    40 & Verification of input             & Creating a program that asks for two inputs and checks if they are the same.  \\
    36 & On calculating an average               & Starter code reads a predefined number of numerical input and outputs their average. It must be fixed so that if no numbers were read, the average is not counted; a specific message is shown instead. \\
    34 & Searching from a phone book         & Creating a function that is given a dictionary (map) as a parameter and that is used for looking for information from the phonebook. The function asks for a phone number (dictionary key) and prints out the owner of the number, if found. Otherwise the function outputs that no owner was found. The function continues asking for a phone number until an empty phone number is provided. \\
    31 & Fixing a bit!                      & Fixing two small errors in a fairly toy program focused on I/O.   \\
    31 & Average distance of long jumps      & Writing a program that reads in values until the user types in a negative number. The program then prints out the average of the inputs or, if no numbers were entered, a specific string indicating that no numbers were provided. \\
    31 & Sum between                       & Creating a two-parameter function that calculates the sum of the numbers between the two given parameters and returns the sum. \\
    28 & Count of entered numbers       & Writing a program that asks for numbers until the user inputs the number zero. The program then outputs the count of the entered numbers.  \\
    28 & Explaining the number                & Creating a single-parameter function that returns a specific string depending on whether the parameter value is negative, positive, or zero.  \\
    23 & First and last name                   & Writing a program that reads in two variables and prints them out in a specific way (``My name is lastname, firstname lastname.'').   \\
    21 & In reverse order & The exercise comes with starter code that reads in a predefined number of values to a list and then prints the last value. The program must be adjusted so that all the values in the list are printed in reverse order. \\
 \bottomrule
\end{tabular}
\normalsize
\end{table*}

\subsection{Context and Data}

Our study is based on data from an open, online introductory programming course organized by Aalto University in Finland. The workload, level of expectations, and breadth differ from normal introductory programming courses at Aalto and in Finland, however. The estimated workload of this course is only 2~ECTS credits (ca. 50 to 60 hours of study) as opposed to the more typical 5~ECTS (ca. 125 to 150h). There are no deadlines, and students can work at their own pace. The course is open to both lifelong learners and Aalto students; we will refer to all participants as ``students.'' 

The course materials are written in Finnish and the programming language is Dart\footnote{\url{https://dart.dev/}}. The topics are typical of classic introductory courses and include standard input and output, variables, conditionals, loops, functions, lists, and maps. 

The course has a bespoke online ebook, which covers the content with a combination of reading materials, worked examples, videos, quizzes, and programming exercises. Students program in their web browser, using a customized DartPad\footnote{\url{https://dartpad.dev}} embedded in the ebook. In addition to DartPad’s default behavior of continuously highlighting syntax errors and running code in the browser, our custom version supports in-browser standard I/O. The exercises are automatically assessed, the platform provides exercise-specific feedback, and there is no limit on the number of submissions. 

A key feature of the platform is the ability to ask for help from teachers. Asking for help is done by clicking a ``Request help'' button. The button resides next to feedback from automated assessment and is at first inactive, but becomes active whenever a student submits an exercise for automated assessment and the solution does not pass the automated tests. Clicking the button opens up a dialog for a help request that gets sent to a queue with the associated exercise details and source code. Course staff responds to the help requests manually. The students also have access to an unofficial chatroom (Slack) with other course participants. 


Our data is from 2022. During the year, there were 4,247 distinct students in the course, who collectively made 120,583 submissions to programming exercises. 831 help requests were submitted. In this article, we focus on the fifteen programming exercises with the most help requests (out of 64 exercises in total). The fifteen exercises, which are summarized in Table~\ref{tbl:programming-exercises-with-most-help-requests}, account for more than 65\% of all the help requests during the year.


For this study, we translated the programming exercise handouts (problem descriptions) to English.
For each of the 15 exercises with the most help requests, we randomly sampled ten, which yielded a body of 150 help requests in total. 

\subsection{Generating LLM Responses to Help Requests}

We generated responses to the help requests with two LLMs: the OpenAI Codex model (\texttt{code-davinci-002}), which is optimized for code, and the GPT-3.5 model (\texttt{gpt-3.5-turbo\footnote{The version released on March 1\textsuperscript{st}, 2023, \url{https://openai.com/blog/introducing-chatgpt-and-whisper-apis}}}) which handles both free-form text and code\footnote{GPT-4 was released on March 14\textsuperscript{th}, 2023 (\url{https://openai.com/research/gpt-4}). While working on this article, we had no access to the GPT-4 API.}. 

We started the analysis with a prompt engineering phase, trying out different types of prompts to find out what produced the most consistent and helpful outputs. We considered the following as potential parts of the prompt: 

\vspace{-1mm}
\begin{enumerate}
    \item The exercise handout
    \item Starter code (where applicable)
    \item The student's code
    \item The help request text written by the student
    \item The model solution
    \item An additional passage of text that describes the context and asks for suggestions
\end{enumerate}
\vspace{-1mm}

During prompt engineering, we observed that the help request texts were unnecessary, as they were generally uninformative beyond indicating that the student was struggling. Another observation was that including the model solution in the prompt often led to a response explaining that solution and increased the chance of the solution being echoed in the response. Moreover, it appeared unnecessary to include trivial starter code (an empty function). 

Of the prompting options that we explored, we deemed the following procedure the best: Begin the prompt with the exercise handout, followed by the student's code and a question. Explain the course context as part of the question. Write the question in the first person (so that the model is likelier to produce output that could be directly given to students). 
Include an explicit request that the model not produce a model solution, corrected code, or automated tests (even though the effect of this request is limited). Include non-trivial starter code and mark it as such in the prompt.

A corresponding prompt template is in Figure~\ref{fig:exercises}. Using this template, we generated responses to our sample of 150 help requests. For \emph{temperature}, a parameter that controls randomness in LLM responses, we used 0, which should yield the most deterministic responses and has been found to work well for feedback in prior work~\cite{leinonen2022using}. To explore the possibility of the model generating the responses in Finnish in addition to English, we created two versions. We thus generated a total of 600 generated help request responses (150 help requests $\times$ 2 languages $\times$ 2 models).

\begin{figure*}
\begin{tcolorbox}
\small
\begin{verbatim}
## Programming exercise handout

<optional starter code>

<handout>

## My code

<student's code>

## Question

I am in an introductory programming course where we use the Dart programming language. I have
been given a programming exercise with the above handout. I have written code given above. My 
code does not work as expected, however. Please provide suggestions on how I could fix my code 
so that it fulfills the requirements in the handout. Do not include a model solution, the 
corrected code, or automated tests in the response.

## Answer
\end{verbatim}
\normalsize
\end{tcolorbox}
\vspace{-2mm}
\caption{Our template for prompting the LLMs. \label{fig:exercises}}
\end{figure*}

\subsection{Classification of Issues in Help Requests}
\label{subsec:method:issues}

The help requests were first analyzed qualitatively, looking for issues in student code. We annotated the source code from the 150 help requests with issues that a teacher would provide feedback on. This was carried out by one of the researchers, who is the teacher responsible for the course, has more than a decade of experience in teaching introductory programming, and has specific experience of answering help requests in this course. We chose to annotate the help requests again instead of using existing answers to these help requests, as the help requests had been previously answered by a pool of teachers and teaching assistants, and we wanted a consistent baseline for the present analysis. 


We then grouped the issues by high-level theme (e.g., logic error, I/O problem) and by sub-theme (e.g., arithmetic, formatting) and determined the themes’ distribution over the exercises. These results are in Section~\ref{sec:results-help-requests}.

 \subsection{Analysis of Help Request Responses}

The LLMs’ responses to help requests were analyzed qualitatively and quantitatively. As Codex often produced surplus content (e.g., new questions and code examples), we cleaned up the data by automatically removing any subsequent content from the responses that repeated the prompt format. 

We focused our analysis on seven aspects, listed below. For each response analyzed, we asked whether it ... 
\vspace{-1mm}
\begin{enumerate}
    \item ... identifies and mentions at least one actual issue?
    \item ... identifies and mentions all actual issues?
    \item ... identifies any non-existent issues?
    \item ... includes duplicate or superfluous content?
    \item ... includes code?
    \item ... includes code that can be considered a model solution?
    \item ... includes any automated tests?
\end{enumerate}
\vspace{-1mm}
\noindent For each of the seven aspects, each LLM response was manually categorized as either `yes' or `no'.

\subsubsection{Comparing Models}
\label{subsubsec:method:comparing}

To gain insight into the relative performance of different LLMs, we conducted an initial analysis on a subset of our data. We randomly chose two help requests for each exercise and analyzed the responses created by GPT-3.5 and Codex with English and Finnish prompts. This step thus involved a total of 120 LLM responses (two help requests $\times$ fifteen exercises $\times$ two models $\times$ two languages), each of which we assessed in terms of the seven questions listed above. The results of this comparison are in Section~\ref{sec:results-llm-comparison}. 

\subsubsection{Analysis of Responses and Issues}
\label{subsubsec:method:extendgpt}

Since the initial analysis suggested that GPT-3.5 clearly outperforms Codex and that its performance is similar in English and Finnish, we focused our subsequent efforts on GPT-3.5’s responses to English prompts. 
After analyzing the remaining 120, we had a total of 150 analyses of English responses from GPT-3.5. 
We combined the classification of issues (Section~\ref{subsec:method:issues} above) with the analysis of the LLM responses, checking how the responses differed for requests that involved different kinds of issues. The results of this analysis are in Section~\ref{sec:results-llm-response-analysis}. 

For further insight, we annotated the LLM responses with free-form notes, noting any phenomena that appeared potentially interesting from a pedagogical point of view; 109 of the 150 responses received at least one annotation. We thematically analyzed these notes; the main results are in Section~\ref{sec:results-llm-response-thematic-analysis}.

\subsection{Ethical Considerations}

The research was conducted in compliance with the local ethical principles and guidelines. To avoid leaking any personal information to third-party services, we manually vetted the inputs that we fed to the LLMs, both during prompt engineering and during the final generation of the responses.

\section{Results}
\label{sec:results}


\subsection{Issues in Help Requests}
\label{sec:results-help-requests}

In 150 help requests, we identified a total of 275 issues, for an average of 1.9 issues per help request. All programs associated with a help request had at least one issue; the maximum was six. 

Six main themes emerged. From most to least common, they are: (1) \emph{logic errors}, present in 108 help requests, 72\%; (2) problems with \emph{input and output}, 51 requests, 34\%; (3) \emph{syntax errors}, 12, 8.0\%; (4) \emph{very incomplete} solutions where the student was far from a working program, 8, 5.3\%; (5) problems with \emph{semicolons}\footnote{The semicolons theme initially emerged as a catch-all category for miscellaneous issues. In the end, however, all these issues involved semicolons immediately after \texttt{if} statements, as in \texttt{if (condition); \{...\}}.}, 4, 2.7\%; and (6) failing to meet \emph{hidden requirements} in automated tests, 3, 2.0\%. Below, we elaborate on the two most common themes.


\subsubsection{Logic errors}

The vast majority of logic errors fell under one of three sub-themes:

\begin{itemize}
    \item \emph{Conditionals} (in 37 requests). E.g., missing conditional, wrong expression in conditional, mistakes in nesting.
    \item \emph{Iteration} (30). E.g., missing iteration, out of bounds errors in loop, incorrect termination.
    \item \emph{Arithmetic} (23). E.g., incrementing a counter incorrectly, summing instead of counting, treating zero as positive.
\end{itemize}

\noindent Other, less common logic errors included misusing function parameters, printing in a function when expected to return a value, misplacing logic, and placing variables outside of functions (leading, e.g., to a sum variable getting incremented over multiple function calls).

\subsubsection{Input and output}

For input/output errors, too, we identified three dominant sub-themes:

\begin{itemize}
    \item \emph{Formatting} of output (25 requests). E.g., completely incorrect formatting, missing information in output, minor extra content in output. This category also includes single-character mistakes in writing and punctuation.
    \item \emph{Unwanted} printouts (24). E.g., debug information printed, completely unexpected output.
    \item \emph{Missing} printouts (10). E.g., failure to produce the specified output when dealing with a corner case.
\end{itemize}

\subsubsection*{Side Note: Exercise Specificity of the Issues}

Different exercises bring about different issues. We explored this briefly, focusing on the most common themes of logic and I/O. As expected, there was considerable variation between exercises. Typically, a single sub-theme was prevalent in a particular exercise (e.g., conditionals in the \emph{Verification of input} exercise; formatting issues in \emph{First and last name}), but there were some exercises with a varied mix of issues.

\subsection{Performance of Different LLMs}
\label{sec:results-llm-comparison}

As described in Section~\ref{subsubsec:method:comparing}, our comparison of the LLMs is based on four LLM--language pairings, with 30 LLM responses analyzed for each pairing, and seven aspects examined for each response. Table~\ref{tbl:brief-comparison-of-llms} summarizes the findings.

\begin{table*}[ht!]
\caption{Comparison of responses by GPT-3.5 and Codex. En = English prompts; Fi = Finnish prompts. \label{tbl:brief-comparison-of-llms}}
\centering
\vspace{-2mm}
\begin{tabular}{lrrrr} 
 \toprule
  Aspect & GPT-3.5 (En) & GPT-3.5 (Fi) & Codex (En) & Codex (Fi) \\ 
 \midrule
    Identifies and mentions at least one actual issue.      & 90\%     & 90\%     & 70\%     & 33\%  \\
    Identifies and mentions all actual issues.              & 57\%  & 53\%  & 13\%  & 17\%  \\
    Identifies non-existent issues.                         & 40\%     & 23\%  & 40\%     & 43\%  \\
    Includes duplicate or superfluous content.              & 0.0\%      & 0.0\%      & 60\%     & 50\%     \\
    Includes code.	                                & 100\%    & 90\%     & 33\%  & 67\%  \\
    Includes a model solution.                     & 67\%  & 70\%     & 13\%  & 40\%     \\
    Includes some automated tests.                 & 0.0\%      & 0.0\%      & 6.7\%   & 10\%     \\
 \bottomrule
\end{tabular}
\end{table*}


The table shows a clear difference in performance between GPT-3.5 and Codex. GPT-3.5 identified and mentioned at least one actual issue in 90\% of the cases in both languages. Codex succeeded 70\% of the time in English, with Finnish performance far behind at 33\%. In terms of identifying \emph{all} of the issues present, GPT-3.5 succeeded approximately 55\% of the time in both languages, whereas Codex’s performance was around a mere 15\%. 

Non-existing issues (false positives) were fairly common in all LLM--language pairings. They were the rarest (23\% of help requests) when GPT-3.5 was prompted in Finnish. Codex was also prone to producing superfluous content.

As for whether the LLMs followed our instructions not to provide sample code or tests, performance was poor across the board. The responses from GPT-3.5 practically always included code, and very often included model-solution-like code. This was less common for Codex, which however did produce automated tests for some of the help requests.

\subsection{Deeper Analysis of GPT-3.5 Responses}
\label{sec:results-llm-response-analysis}

\subsubsection{Results from an Extended Dataset}

As described in Section~\ref{subsubsec:method:extendgpt}, we proceeded by analyzing all the 150 responses produced by GPT-3.5 with the English prompts. Table~\ref{tbl:gpt-35-en-help-request-responses} summarizes the findings, which are similar to those we obtained for GPT-3.5 with the smaller dataset and reported above.

\begin{table}[ht!]
\caption{The performance of GPT-3.5 on 150 help requests, prompted in English. \label{tbl:gpt-35-en-help-request-responses}}
\centering
\vspace{-2mm}
\begin{tabular}{lr} 
 \toprule
    Aspect & Proportion \\ 
 \midrule
    Identifies and mentions at least one actual issue.      & 82\% \\
    Identifies and mentions all actual issues.              & 55\% \\
    Identifies non-existent issues.                         & 48\% \\
    Includes duplicate or superfluous content.              & 0.0\%  \\
    Includes code.	                                & 99\% \\
    Includes a model solution.                     & 65\%  \\
    Includes some automated tests.                 & 0.0\%   \\
 \bottomrule
\end{tabular}
\end{table}

For 123 help requests out of 150, GPT-3.5 correctly identified and mentioned at least one actual issue; for 82 of those, it identified and mentioned all actual issues. The LLM identified non-existing issues in 72 help requests. 

Even when it did not mention the actual issues, GPT-3.5 often generated model-solution-like code. Almost every response included code, and the code was model-solution quality in roughly two responses out of three. 

Given that we had grouped the issues in student code (Section~\ref{sec:results-help-requests} above), it was easy to break down the GPT-3.5 analysis by issue type, so we did that. Table~\ref{tbl:gpt-35-mapping-responses-and-issues} summarizes. 
  
\begin{table*}[ht!]
\caption{The performance of GPT-3.5, in English, on the 150 help requests, split by issue type.\label{tbl:gpt-35-mapping-responses-and-issues}}
\centering
\vspace{-2mm}
\begin{tabular}{ll|ccl}
 \toprule
         &  & \multicolumn{3}{l}{GPT-3.5 identifies and mentions issues:} \\
     Theme of issue & Sub-theme    & One (or more) & All & Non-existent(s)                 \\
    \midrule
    Logic error & Conditionals (n=37)  & 86\%                                            & 35\%                                      & 49\%                           \\
    & Iteration   (n=30)  & 97\%                                            & 73\%                                      & 40\%                           \\
    & Arithmetic  (n=23) & 91\%                                             & 57\%                                      & 35\%                           \\
    \midrule
    Input / output & Formatting  (n=25) & 72\%                                             & 44\%                                      & 52\%                           \\
    & Unwanted    (n=24) & 75\%                                             & 54\%                                      & 63\%                           \\
    & Missing     (n=10) & 70\%                                             & 50\%                                      & 50\%                           \\
    \midrule
    Other & Syntax      (n=12) & 92\%                                             & 50\%                                       & 50\%                           \\
    & Very incomplete (n=8)  & 100\%                                            & 63\%                                       & 13\%                           \\
    & Semicolons  (n=4)  & 100\%                                            & 100\%                                      & 25\%                           \\
    & Hidden req     (n=3)  & 33\%                                             & 0.0\%                                        & 67\%                           \\
 \bottomrule
\end{tabular}
\end{table*}

\emph{Note:} In many cases, a help request had more than one issue (1.9 on average), and our analysis does not account for whether the help request responses addressed a specific issue type. 

Consider the logic errors theme in Table~\ref{tbl:gpt-35-mapping-responses-and-issues}. When issues related to Conditionals are present, the LLM addresses all the issues in 35\% cases; with Iteration issues are present, the same proportion is 73\%; and when Arithmetic issues are present, it is 57\%. For the input/output theme, the proportions are somewhat lower: 44\%, 54\%, and 50\% for formatting issues, unwanted outputs, and missing outputs, respectively. 


\subsubsection{Exercise-Specific Results}

We briefly looked into how specific exercises interplay with the performance of GPT-3.5. Table~\ref{tbl:finding-and-addressing-issues-in-exercises} summarizes the results of this supplementary analysis.
 
\begin{table*}[ht!]
\caption{GPT-3.5 performance on help requests related to specific programming exercises. Each row describes GPT’s behavior on requests related to that exercise. The figures are out of ten, as we sampled ten help requests per exercise. \label{tbl:finding-and-addressing-issues-in-exercises}}
\centering
\vspace{-2mm}
\begin{tabular}{l | ccl }
 \toprule
          & \multicolumn{3}{l}{GPT-3.5 identifies and mentions issues:} \\ 
     Exercise    & One (or more) & All & Non-existent(s)            \\ 
 \midrule
    Difference between two numbers & 6  & 1  & 9 \\ 
    Asking for a password          & 10 & 8  & 4 \\ 
    Average of entered numbers     & 10 & 6  & 5 \\ 
    Counting positive numbers      & 10 & 7  & 2 \\ 
    Authentication                 & 9  & 6  & 5 \\ 
    Verification of input          & 8  & 4  & 5 \\ 
    On calculating an average      & 8  & 6  & 4 \\ 
    Searching from a phone book    & 8  & 4  & 5 \\ 
    Fixing a bit!                  & 7  & 5  & 5 \\ 
    Average distance of long jumps & 8  & 6  & 4 \\ 
    Sum between                    & 9  & 6  & 5 \\ 
    Count of entered numbers       & 8  & 7  & 4 \\ 
    Explaining the number          & 5  & 1  & 9 \\ 
    First and last name            & 9  & 9  & 1 \\ 
    In reverse order               & 10 & 10 & 4 \\ 
 \bottomrule
\end{tabular}
\end{table*}

As shown in the table, there are exercise-specific differences in the extent to which the responses address the issues; there is no obvious pattern, however. In the worst-case scenario, the responses address all of the issues in only one response out of the ten that we sampled; in the best case, all issues are addressed in ten of ten responses. Even in the latter case, however, four of the ten responses featured false positives. To illustrate, here is some student code. 

\small
\begin{lstlisting}
for(var i = list.length; i>= 0; i--) {
   var value = list[i];
   
   print('$value');
   }
\end{lstlisting}
\normalsize

The variable \texttt{i} is inappropriately initialized: its initial value should be \texttt{list.length - 1}, which GPT-3.5’s response correctly identified and mentioned. However, the response also suggested an `imaginary' issue: \emph{``Also, you have an extra closing curly brace at the end of the code block. Remove that to avoid a syntax error.''}


\subsection{Further Insights: Thematic Analysis of Researcher Notes}
\label{sec:results-llm-response-thematic-analysis}

\subsubsection{Language and Tone}

As described in Section~\ref{subsubsec:method:extendgpt}, we collected and thematically analyzed free-form notes about the 150 responses from GPT-3.5. 

All of the LLM responses were phrased as actual attempts to help. A large majority had a confident tone; this was the case even where the advice was completely wrong. Fewer than ten of the responses had a somewhat non-confident tone, employing phrases such as \emph{``the issue might be,''} \emph{``the code seems,''} or \emph{``the issue seems.''} 
Of the 150 responses, 27 encouraged the student with phrases such as \emph{``you are close to the solution,''} \emph{``you are on the right track,''} \emph{``your code looks good,''} \emph{``your code is mostly correct but ...''} We observed no negativity in any of the responses. 

There was some variation in terms of agency. 78 responses attributed actions to the student: what they did or should do, as in \emph{``when you initialize,''} \emph{``you need to,''} or \emph{``you can.''} Nineteen responses implied a shared activity or a passive ``we,'' as in \emph{``we need to,''} \emph{``we can,''} or \emph{``we should.''} In nine responses, the LLM itself was given agency, as in \emph{``I would.''}

At least twenty of the 150 responses featured a discrepancy between the explanation and the code in the response. For example, one response emphasized how a \texttt{readlineSync} function might return a line break, even though the actual issue (deliberately inserted in starter code) involves letter case when calling Dart’s \texttt{readLineSync}; however, the model-solution-like code in the LLM’s response \emph{was} correctly lettered. In another example, the LLM suggested moving a \texttt{return} statement to the end of a function---but did not actually implement this advice in the code that it provided. 


Three more topics surfaced from this analysis that we felt were of potential significance for instructional design. First, and as also highlighted in Table~\ref{tbl:gpt-35-en-help-request-responses}, almost all of the responses included code and most effectively provided model solutions. Second, in 17 of the responses (over 10\%), the LLM suggested adding functionality that had not been covered in the course and was not in the course plan; these suggestions included error handling, null safety features of Dart, and specific library functions for list processing. Third, and again related to the model solutions, for all the six help requests where we classified the student’s code as very incomplete (i.e., far from the actual solution), the response was pedagogically unsuitable in that it did not focus on what would be relevant to the student at such a stage. The following scenario outlines one instance of this. 

Example: A student asks for help on \emph{Average distance of long jumps} with the following code. 

\begin{figure}[h!]
\begin{lstlisting}[breaklines=true]
  print('Type in jump lengths. Negative input stops reading.');
int.parse(stdin.readLineSync());
  while (true)}
\end{lstlisting}
\end{figure}

The student’s code has a syntax error and would have an an infinite loop if the syntax were to be corrected. The LLM responds in very encouraging tones, but not as a teacher might prefer:

\small
\begin{quote}
\emph{You are on the right track by using a while loop to continuously ask the user for input until a negative value is entered. However, there are a few issues with your code.}
\end{quote}\begin{quote}
\emph{Firstly, you are not storing the input values in a variable. You need to read the user input with 
\texttt{stdin.readLineSync()} and store it in a variable. You can then convert the input to an integer 
with `int.parse()`.}
\end{quote}\begin{quote}
\emph{Secondly, you need to check if the input value is negative and break out of the loop if it is. You can do this with an if statement.}
\end{quote}\begin{quote}
\emph{Lastly, you need to calculate the average length of jumps and print it to the console. You can keep track of the total sum of jumps and the number of jumps entered in two separate variables. You can then divide the total sum by the number of jumps to get the average length. If no jumps were entered, you should print "No jumps." instead.}
\end{quote}\begin{quote}
\emph{Here is some modified code that should work:} [model solution omitted for brevity]
\end{quote}\begin{quote}
\emph{Note that I have added a \texttt{!} after \texttt{stdin.readLineSync()}. This is called null assertion 
operator and tells the Dart compiler that the value returned by \texttt{stdin.readLineSync()} will 
never be null.}
\end{quote}
\normalsize

\noindent Our notes also recorded whenever a response was good enough to give to students as is. This was rare, especially since most responses included model solutions: fewer than ten of the 150 responses were considered presentable without editing, assuming that the objective was to help students in a pedagogically sensible way. This number would go up if a script were to prune out code from the responses, but we did not explore this further.




  
  




\section{Discussion}
\label{sec:discussion}

Subsections 5.1--5.4 below discuss our main research interest: answering help requests with LLMs. Subsection 5.5 provides additional observations on student help seeking, issues in student code, and contextual factors. 

\subsection{LLM Performance on Help Requests}

Both large language models were able to identify some issues in the help requests, but GPT-3.5 was considerably more accurate than Codex. Overall, GPT-3.5 might be described as quite effective at issue-hunting, but it is far from reliable in terms of finding all the issues, and false positives are common as well.

Our main analysis focused on the LLM responses that were produced with GPT-3.5 in English. We observed that the model identified and mentioned at least one actual issue in 82\% of the help requests; all were identified and mentioned in 55\% of the cases.  `Mentioning' an issue, in our sense, implies also suggesting how to fix the issue; this is more than most feedback systems for programming exercises do, as they tend to focus on identifying student mistakes~\cite{keuning2018systematic}. 

A significant limitation to the quality of GPT-3.5’s responses is that 48\% of them reported on issues that did not actually exist in the student’s code. Such responses may lead students down a ``debugging rabbit hole''~\cite{vaithilingam2022expectation} as the student tries to fix non-existent issues while remaining oblivious to actual ones. This phenomenon of LLMs often ``hallucinating'' false information has been highlighted by many~\cite{ji2022survey}. The confident tone of the LLM responses---we observed just a handful of responses in less-than-confident tones---may exacerbate the problem.

In our brief exploration of non-English prompts, GPT-3.5 performed similarly in Finnish as in English in terms of the LLM’s ability to identify issues in code. The Finnish in the LLM’s responses was also in general understandable and could have been shown to students, as far as language quality was concerned. This suggests that responses from large language models are potentially viable in non-English-speaking classrooms.

\subsection{Pedagogical Quality of LLM Feedback}

\subsubsection{The Problem of Model Solutions}

Even though we explicitly prompted GPT-3.5 not to produce model solutions, corrected code, or automated tests, almost every response did include code, and two responses out of three essentially provided a model solution for the exercise. Similar phenomena have been acknowledged as a limitation of LLMs, and recent research efforts have improved LLMs’ ability to follow  instructions~\cite{ouyang2022training}; this has been claimed as an improvement in the recently released GPT-4, for example~\cite{openaigpt4}. The instant provision of model solutions poses some obvious problems from a pedagogical point of view. Nevertheless, we note that there are cases where model solutions are useful: for example, model solutions have been deliberately provided to students in some prior research~\cite{nygren2019experimenting,nygren2019non}, and they are also often provided in automated learning environments that are not focused on grading~\cite{jeuring_towards_2022}. It would also be possible to create a parser for LLM responses that strips away code before relaying the response to students. 

Even if LLM responses to help requests are not directly sent to students, they might be used to help teachers respond to requests. One option is to employ an LLM to create template responses, which are then edited by teachers. This might also be explored in the context of programming error messages~\cite{becker2019compiler,denny2021designing,leinonen2022using} as well as in feedback systems that group similar submissions together so that feedback may be provided to many students at once~\cite{glassman2015overcode,head2017writing,nguyen2014codewebs,koivisto2022evaluating}.

\subsubsection{The Problem of Effective Feedback}

Some of the LLM responses included words of encouragement. The LLM might state, for example, that \emph{``you are on the right path''} or vaguely praise the student. Positivity can certainly be desirable in feedback, but it is challenging to provide just the right kind of supportive feedback that takes the student's level, context, and other factors into account~\cite{ott2016translating}. Praising on easy tasks may lead students simply to dismiss the feedback; at worst, it may implicitly suggest that the student lacks ability~\cite{borich2005educational} and demotivate the student. 

Instructional guidance should attend to the student’s current level of domain knowledge; a mismatch will result in poorer learning outcomes~\cite{kalyuga2007expertise}. Although the LLM responses sought to address the technical issues and at times provided positive feedback, we saw little indication of the feedback being adjusted to the (beginner) level of the programming exercises being solved or to the context (the introductory course that we mentioned in the prompt). A handful of the LLM responses included suggestions that were well beyond the scope of an introductory programming course. In this work, we did not even attempt to describe student-specific levels of prior knowledge to the LLMs.

Future work should explore the creation of large language models that are `aware' of students’ 
evolving prior knowledge and competence in programming. Such LLMs might then generate feedback messages that match the level of the particular student. One potential direction for this work is to track the time that the student has spent on a task, which has been observed as one of the indicators of programming exercise difficulty~\cite{ihantola2014automatically} and which correlates with performance~\cite{leinonen2017comparison,leinonen2022time}; the LLM could be fine-tuned to take task difficulty into consideration. Fine-tuning an LLM to match specific course progressions is also a possibility. Moreover, it might be fruitful to distinguish  feedback on progress from suggestions about fixing specific issues. Here, approaches such as adaptive immediate feedback~\cite{marwan2020adaptive} and personalized progress feedback~\cite{leppanen2022piloting} could be meaningful directions.

\subsection{The Need to Comprehend Code}

The proliferation of LLMs and their inevitable use by both novices and professional programmers lends further emphasis to program comprehension as a key skill. Programmers need to understand  code and learn to debug code created by others---where ``others'' now includes LLMs. 
Although LLMs are a partial cause of the situation, they may also be part of the solution. Even with the deficiencies that LLMs now have (e.g., inaccuracy and confident hallucinations), they could potentially be taken into use in programming courses as long as the issues are acknowledged. For example, if it is clear enough to students that the code created by LLMs is often faulty, a novel type of learning activity might involve students evaluate LLM-created code to spot issues and improve the code, with the aim of teaching code comprehension, debugging, and refactoring in the process. In addition to potentially being educational for the students, such activities could be used to further tune LLM by giving it the improved code as feedback.


\subsection{On the Evolution of LLMs}

The evolution of large language models has been extremely rapid recently, and only seems to accelerate. We conducted our analysis in March 2023, at a time when GPT-3.5-turbo from March 1\textsuperscript{st} was the most recent model readily available. At the time of writing, however, the most recent model is GPT-4, which reportedly performs better on most tasks.

Our results suggest that this evolution is also visible in performance on the task we are interested in, responding to student help requests. Comparing the older Codex LLM to the newer GPT-3.5, we found that GPT-3.5 outperformed Codex. This raises interesting questions about how long the results of LLM performance studies are valid. For example, much of the prior work in CER has employed LLMs that are already `ancient.'

The rapid evolution can be troublesome for research replication and for the integration of LLMs into teaching. For example, on March~21\textsuperscript{st}, 2023, OpenAI announced that support for the Codex API will be discontinued within days. This renders our results on Codex performance nearly impossible to replicate. Such developments highlight the importance of truly open LLMs that can be run locally without relying on third-party APIs.


\subsection{Additional Observations}

\subsubsection{Student Help-Seeking and Related Research}

In our target course, each help request is linked to a specific exercise submission. However, of all the submissions in the course, only a tiny fraction have associated help requests. During 2022, we got 120,583 submissions but only 831 (0.7\%) of them had a help request. We checked whether this could be due to students' mostly submitting correct solutions, but that was not the case: only 56,855 submissions (47\%) passed all the tests. This means that the students asked for help with only 1.3\% of the 63,728 failed submissions.

Asking for help is difficult~\cite{butler1998determinants,karabenick2004perceived,ryan1998some,ryan2001avoiding,seamark2018barriers}, but even so, the low proportion of help requests underlines that nearly all failing submissions are such where students do not explicitly request for help. This raises a question related to research based on students’ code submissions and errors therein:
the vast majority of prior research has not explicitly collected information on whether students want help, so some earlier findings about student `failure' may in fact be related to students employing the submission system as a feedback mechanism, not necessarily needing help but simply checking whether they are on the right path. If so, prior research such as the reported mismatch between educators' beliefs about students' mistakes and logged data about mistakes~\cite{brown2014investigating} might be explained in part by students asking for help from educators only when they really need help, which might differ from how they employ automated systems.

%

We acknowledge that the help request functionality in this course is not something that every student is eager to use. Some students will have needed help but decided not to ask for it. Prior research on a help request platform for programming noted that only one third of the students who open up a help request dialog end up writing a help request, and that even the prompts used in the help request dialog can influence whether a help request gets sent~\cite{sarsa2022help}. Platform functionality aside, students may seek help from many sources, such as their peers or online services~\cite{liao2019behaviors,muller2020exploring,nelimarkka2018social,loksa2016programming}---now from public LLMs as well. Instead of seeking help, students may also resort to plagiarism~\cite{hellas2017plagiarism} or simply drop out~\cite{onah2014dropout}. 

Future research should seek to detect when students need help in order to provide timely feedback~\cite{jeuring_towards_2022,mao2019one,leinonen2022comparison}. That research might be informed by prior work, which has highlighted that data collected from the programming process encodes information about students' struggles~\cite{mao2019one,jadud2006methods,watson2013predicting,becker2016new,petersen2015exploration,carter2015normalized,ahadi2015exploring,leinonen2022time}. Including such process data in help requests has unexplored potential that might be fulfilled through dedicated LLMs, for example.

\subsubsection{Context-Dependent Issues in Student Code}

Many of the student programs had multiple issues, and some types of issues were more frequent than others. This is unsurprising and in line with prior research on logic and syntax errors~\cite{jadud2005first, denny2012all, vihavainen2014novices,dy2010detector,mccall2014meaningful,denny2012all,brown2017novice,smith2019error,mccall2019new,altadmri201537}. Like Ettles et al.~\cite{ettles2018common}, and again unsurprisingly, we observed that the distribution of issues depended on the exercise.  

Many of the help requests involved input and output (with the theme present in 34\% of the requests). These issues were especially common very early on in the course, when students were practicing I/O. Upon further reflection, some of the issues are in part explained by worked examples in the course materials: for example, in one exercise, many students had incorrect formatting apparently because they had copied the output format not from the exercise handout but from a similar-looking worked example that immediately preceded the exercise. Such struggles with I/O reduced later on in the course, perhaps suggesting that students learned to pay closer attention to the handouts.

In some cases, it was evident that a student had correctly interpreted what they were supposed to achieve overall, but had omitted a part of the solution, perhaps simply by overlooking a requirement or because they were testing the program with specific inputs. This was present especially in the more complex `Rainfall-problem-like' exercises of the course, which are known to be challenging for some novices~\cite{soloway1982what,seppala2015we}. It is possible that students were overloaded by the complexity of these problems and, upon reaching a solution that works in a specific case, failed to attend to identify the rest of the requirements. Pedagogical approaches that emphasize full understanding of the problem handout~\cite{hilton2019translation,prather2019first}, or even brief quizzes requiring the student to read the problem in more detail, could be beneficial~\cite{vihavainen2015benefits}. 

In comparison with some prior studies that have reported data on syntax errors, our sample had relatively few: syntax errors featured in only 8.0\% of the help requests. This may be in part because we were dealing with code that students had chosen to submit for marking. These syntax errors were often jointly present with other types of issues. 

Anecdotally, our dataset had several instances of the dreaded semicolon-in-conditional: \texttt{if (foo); \{ ... \}}. This issue has been observed to take significant amounts of time for students to fix~\cite{altadmri201537}.

    





\section{Limitations}

The particular features of our context limit the generalizability of our findings. The course uses a relatively recent programming language (Dart), which the LLMs will not have `seen' as much as some other programming languages. 
Moreover, the course is fully online, and its scope and student cohort are different from many typical introductory programming courses at the university level; the issues that students request help with might thus differ from other courses. 

Relatedly, only a minority of code submissions had an associated help request. There are many possible explanations for this. For example, students may rely on other sources for help, such as the course materials or internet searches. It is also possible that the students who use the built-in help request functionality differ from the general course population.

A major limitation is that there are already newer LLMs than the ones we used. As we submitted this article for review, GPT-4 represented the state of the art, but we did not have programmatic access to it. Anecdotally, after receiving access to it, we have observed that GPT-4 outperforms GPT-3.5, to an extent, but does not fully eliminate the challenges highlighted in this article.

Our qualitative analysis employed a single coder, which is a threat to reliability.

In the present study, we relied on a single request to the model. However, LLM-based applications such as ChatGPT enable ongoing discussions with an evolving context, which is something we did not explore. In our future work, we are interested in studying a more conversational approach to providing feedback to students, which might more closely match the dialogue between a teaching assistant and a student, or a student and an LLM. One way to potentially achieve this could be to fine-tune the LLM to avoid giving correct answers and instead help the student arrive at the solution.



\section{Conclusions}
\label{sec:conclusion}

In this study, we have taken a look at how two large language models---OpenAI Codex and GPT-3.5---perform in analyzing code that accompanies students' help requests in a particular online course on programming basics.

Overall, we find that the LLMs’ responses are usually sensible and potentially helpful (RQ2). GPT-3.5 in particular was good at identifying issues in student code. However, these LLMs cannot be counted on to identify all the issues present in a piece of code; they are also liable to report on `imaginary' non-issues and to mislead students. At least in this context and with these LLMs, output formatting surfaced as a difficult topic for the LLMs. Although the LLMs appear to perform best in English, responses in a fairly uncommon non-English language were not far behind in quality.

We see LLMs as a potentially excellent supplement for programming teachers and teaching assistants, available at scale to serve the ever-increasing numbers of programming students. Not as a \emph{replacement} for teachers, however. If we dismiss for a moment the risks of anthropomorphisms, we may describe an LLM as a beginner programmer's quick-thinking, often helpful but unreliable tutor friend, who has plenty of experience with code but less of an understanding of good pedagogy, and who has a penchant for blurting out model solutions even when you directly ask them not to.

Our study also presented us with a window into students’ help-seeking behavior (RQ1). We found that: students infrequently asked for help even when their code submissions were failing; most issues involved program logic or input/output; and the I/O issues might stem from worked examples in the course materials. These findings, too, are specific to the studied context and their generalizability remains to be determined. 

The capabilities and availability of large language models mean that they will be a part of programming education in the future---they are already a part of it today. Computing educators and computing education researchers must find out how to employ these tools productively and to avoid their pitfalls. Future programming students might benefit not only from generic LLMs such as the ones we studied but also from custom LLMs designed and taught to serve the needs of student programmers. We hope that our research is a step towards that goal.

\begin{acks}
We are grateful for the grant from the Ulla Tuominen Foundation to the second author.
\end{acks}

\bibliographystyle{ACM-Reference-Format}
\bibliography{sample-base}

\end{document}